\documentclass[pra,aps,twocolumn]{revtex4}
\usepackage{graphicx,amsmath}
\usepackage{bbm}
\usepackage{txfonts}
\usepackage{hyperref}
\newcommand{\rp}[1]{(\ref{#1})}

\newcommand{\abs}[1]{\left|{#1}\right|}

\newcommand{\av}[1]{\left\langle #1 \right\rangle}

\newcommand{\br}[1]{\langle #1|}

\newcommand{\ke}[1]{|#1\rangle}

\newcommand{\kb}[2]{\ke{#1}\br{#2}}

\newcommand{\al}[1]{^{(#1)}}
\newcommand{\da}{^\dagger}

\newcommand{\pt}[1]{\left( #1 \right)}
\newcommand{\pq}[1]{\left[ #1 \right]}
\newcommand{\pg}[1]{\left\{ #1 \right\}}

\newcommand{\bs}[1]{\boldsymbol #1}

\newcommand{\lpq}[1]{\left[ #1 \right.}
\newcommand{\lpg}[1]{\left\{ #1 \right.}

\newcommand{\rpq}[1]{\left. #1 \right]}
\newcommand{\rpg}[1]{\left. #1 \right\}}
\newcommand{\ee}{{\rm e}}
\newcommand{\ii}{{\rm i}}

\newcommand{\id}{\mathbbm{1}}

\newcommand{\nn}{{\nonumber}}

\newcommand{\mat}[2]{ 
                      \begin{array}{#1}
                       #2
                       \end{array}  }

\newcommand{\vbe}{{\bs \beta}}

\newcommand{\vb}{{\bf b}}

\newcommand{\vv}{{\bf v}}

\newcommand{\DD}{{\cal D}}

\newcommand{\EE}{{\cal E}}

\newcommand{\LL}{{\cal L}}
\newcommand{\MM}{{\cal M}}

\newcommand{\WW}{{\cal W}}

\begin{document}

\title{Steady-state nested entanglement structures in harmonic chains with single-site squeezing manipulation}

\author{Stefano~Zippilli, Jie~Li, David~Vitali}
%\author{S. Z.}
\affiliation{School of Science and Technology, Physics Division, University of Camerino, via Madonna delle Carceri, 9, I-62032 Camerino (MC), Italy, and INFN, Sezione di Perugia, I-06123 Perugia, Italy}

\begin{abstract}
We show that a squeezed bath, that acts on the central element of a harmonic chain, can drive the whole system to a steady state featuring a series of nested entangled pairs of oscillators that, ideally, covers the whole chain regardless of its size. We study how to realize this effect in various physical implementations, including optomechanical and superconducting devices, 
using currently available technologies. In these cases no squeezed fields are actually needed, and the squeezed bath is, instead, simulated by quantum reservoir engineering with bichromatic drives.
\end{abstract}

\maketitle

\section{Introduction}

Quantum technologies and engineering exploit entanglement to boost the efficiency of quantum devices.
In spite of the many small-scale demonstrations of engineered quantum dynamics, the efficient control over large ensembles of quantum systems, for realistic realizations of scalable quantum technology applications, still seems to require significant effort from both the experimental and the theoretical side~\cite{Georgescu,Aolita}.
Among the many strategies envisaged and pursued for quantum manipulations, 
reservoir engineering is particularly intriguing because it makes use of irreversible/dissipative processes to achieve the designed result at the steady state of controlled open quantum dynamics~\cite{Poyatos,Carvalho,Pielawa,Barreiro,Krauter,Shankar,Lin13,Kienzler,Wollman,Pirkkalainen}.
These techniques have been also applied to the manipulation of arrays, proving to  be in principle applicable to a scalable architecture for quantum computation and information~\cite{Diehl,Kraus08,Verstraete,Cho}. Nevertheless, in the schemes proposed so far, many control fields are typically needed to engineer the dynamics of the many elements, and the complexity of the problem increases fast with their number, making these strategies of difficult implementation with present day technologies.

\begin{figure}[!t]
\begin{center}
\includegraphics[width=8cm]{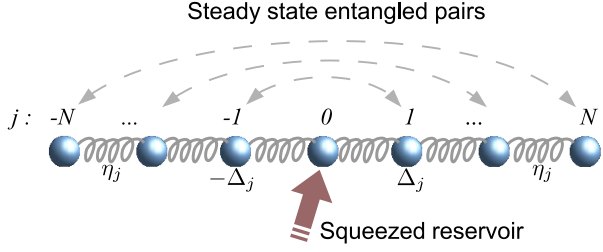}
\end{center}
\caption{(Color online)
An array of linearly coupled harmonic oscillators is locally driven by a squeezed reservoir. The elements in the array are labeled by the indices $j\in\pq{-N,N}$, with $j=0$ indicating the driven central oscillator. The steady-state entangled pairs are marked by dashed arrows.
}\label{fig0}
\end{figure}

On the other hand it has been shown that in certain cases it suffices to have access only to a limited part  of an ensemble of interacting elements to control the whole dynamics~\cite{Zippilli13,Zippilli14,Ma14,Wendenbaum}. In this context, here we propose a new scheme for the preparation of entangled states in a chain of harmonic oscillators with nearest-neighbour coupling, by reservoir engineering, with active control over only a single element of the chain.
Specifically we show that a harmonic chain can approach an entangled steady state when only the central element of the chain is driven by a single-mode squeezed reservoir. 
In particular the oscillators of the chain get entangled in pairs. Each pair is formed by the two oscillators at equal distance, on the left and on the right, from the central one, as sketched in Fig.~\ref{fig0}.
Ideally, all the pairs are entangled with the same degree of entanglement (which is directly related to the degree of squeezing of the bath) irrespective of the size of the chain. 
This is the result of the interplay between localized irreversible dynamics and coherent interactions under specific symmetry conditions. 
Similar results, concerning pairs of entangled systems belonging to two separated chains have been already discussed in Refs.~\cite{Zippilli13,Zippilli14}, which extended and generalized previous works on the transfer of entanglement to distant quantum objects~\cite{Kraus04,Paternostro04,Adesso10}. However,
we remark that, differently from the scheme discussed in Refs.~\cite{Zippilli13,Zippilli14}, where the squeezed bath consists of two entangled modes each of which drives two independent arrays, here we consider a single squeezed mode which interacts with a single element of the chain. 
Hence the idea of entanglement transfer is absent and, here, no spatially separated entangled fields are required.

We study the performance of this protocol in the case of optomechanical and circuit-QED systems where the harmonic chain is realized by, respectively, mechanical and coplanar-waveguide resonators. In these implementations no squeezed resources are actually needed. In fact, in both cases an effective squeezed reservoir is realized by coupling the central resonator to an additional system 
driven by a two-tone field, which in appropriate parameter regimes can simulate a squeezed  reservoir in the spirit of reservoir engineering~\cite{Cirac93,Rabl04,Mari09,Kronwald13,Kronwald14,Porras12}. In the first case the additional system is an optical cavity while in the second a superconducting flux qubit.
These results may find application in the context of quantum information as efficient strategies to generate entanglement between remote, non-directly interacting elements of a quantum processor.

The article is organized as follows. In Sec.~\ref{ideal} we describe the general model and the symmetry of the chains that are required for the observation of the steady-state entangled pairs. The properties of the steady state are then studied in detail in Sec.~\ref{steady state}. Possible physical implementations are identified in Sec.~\ref{physcalSystems}. Finally, Sec.~\ref{conclusions} is devoted to outlooks and concluding remarks.

\section{General Model: Chain of harmonic oscillators with a localized squeezed reservoir}
\label{ideal}

In this Section we analyze the ideal model that is central to this work,
and we identify the necessary symmetries the model has to fulfill in order to sustain the steady state entangled pairs. We consider a chain of harmonic oscillators with nearest-neighbour coupling and with a localized squeezed reservoir which affects only the central oscillator.
Specifically, we consider a chain of $2N+1$ linearly coupled harmonic oscillators, with coupling strengths $\eta_j$, as depicted in Fig.~\ref{fig0}, and with creation and annihilation operators $b_j\da$ and $b_j$, for $j\in\pg{-N,\cdots 0,\cdots N}$,
where the index $j=0$ corresponds to the central oscillator while negative and positive indices indicate, respectively, the oscillators to the left and to the right of the central one in such a way that  $j=-N$ and $j=N$ indicate, respectively, the leftmost and the rightmost oscillator. 
We assume that the coupling strengths follow a symmetric pattern with respect to the central oscillator such that the coupling between the oscillators $j$ and $j-1$ is equal to that between $-j$ and $-j+1$. The frequencies of the oscillators, $\omega_j$,  follow instead an antisymmetric pattern with respect to the central oscillator. In detail, pairs of oscillators with indices $j$ and $-j$ (for $j\in\pg{1,\cdots N}$) are detuned from the central oscillator (the frequency of which is $\omega_0$) by opposite quantities, 
that is $\omega_j-\omega_0=\omega_0-\omega_{-j}=\Delta_j$. 
The effects of deviations from this symmetric situation are discussed in the next section.
The corresponding Hamiltonian, in the interaction picture with respect to $H_0=\hbar\,\omega_0\sum_{j=-N}^N b_j\da\, b_j$, is 
given by 
\begin{eqnarray}\label{Hchain}
&&H_{chain}=\hbar\sum_{j=1}^N\Delta_j\pt{b_j\da\ b_j-b_{-j}\da\ b_{-j}}
\\&&
\hspace{.6cm}+\hbar\sum_{j=1}^N \eta_j\pt{b_{j-1}\ b_j\da+b_{j-1}\da\ b_j+b_{-j+1}\ b_{-j}\da+b_{-j+1}\da\ b_{-j}}\ .
\nn
\end{eqnarray}
As shown in detail in the next section, by virtue of these symmetry conditions the chain is invariant under the effect of specific Bogoliubov transformations. The corresponding Bogoliubov modes are combinations of pairs of oscillators, and are the modes that are actually stabilized by the effect of the squeezed bath, that drives, with rate $\Gamma$, only the central oscillator. Its  
%
%{\it The central oscillator is driven with rate $\Gamma$ by a single-mode squeezed reservoir (as for example, the output field of a degenerate parametric oscillator, realizing a system similar to the one studied in Ref.~\cite{Zippilli14}), whose}
%
effect on the system density matrix $\rho$ can be described in terms of a Lindblad operator, $\LL_{squeez}\al{\Gamma,\bar n,\bar m}$, of the form~\cite{GardinerZoller} 
\begin{eqnarray}\label{LLsqueez}
\LL_{squeez}\al{\Gamma,\bar n,\bar m}\rho&=&\Gamma\lpg{
(\bar n+1)\,\DD\pq{b_0,b_0\da}+\bar n\,\DD\pq{b_0\da,b_0}
}\nn\\&&\rpg{
-\bar m\,
\DD\pq{b_0,b_0}-\bar m^*\,\DD\pq{b_0\da,b_0\da}
}\rho
\end{eqnarray}
with
\begin{eqnarray}\label{DD}
\DD\pq{x,y}\rho=2\ x\ \rho\ y-y\ x\ \rho-\rho\ y\ x\ .
\end{eqnarray}
This model describes, for example, the effect of the output field of a degenerate parametric oscillator (similar to the system studied in Ref.~\cite{Zippilli14}).
The statistical properties of the squeezed reservoir are determined by the parameters $\bar n$, which accounts for the number of excitations, and $\bar m$, which accounts for the correlations. In general $\abs{\bar m}$ is bounded by the value
$\abs{\bar m}\leq\sqrt{\bar n(\bar n+1)}$, with the equality corresponding to a pure state of the reservoir. When $\abs{\bar m}<\sqrt{\bar n(\bar n+1)}$, instead, the reservoir is in a squeezed thermal state. Specifically, the quadratures of the reservoir are squeezed below the vacuum level only when $\abs{\bar m}>\bar n$, 
and the variance of the maximum squeezed quadrature (relative to the vacuum noise level) is 
\begin{eqnarray}\label{S}
S=2\,\bar n+1-2\abs{\bar m}.
\end{eqnarray}
Finally in our description we also take into account thermal dissipation at rates $\gamma_j$ in a bath  with $n_{Tj}$ thermal excitations, which can be described by a Lindblad operator of the form 
\begin{eqnarray}\label{LLdiss}
\LL_{diss}\rho&=&\sum_{j=-N}^N\gamma_j\pg{
\pt{n_{Tj}+1}\DD\pq{b_j,b_j\da}+n_{Tj}\DD\pq{b_j\da,b_j}
}\rho\ .
\nn\\
\end{eqnarray}
In summary we study a system described by the following master equation
\begin{eqnarray}\label{Meq0}
\dot\rho&=&-\frac{\ii}{\hbar}\pq{H_{chain},\rho}+
\LL_{squeez}\al{\Gamma,\bar n,\bar m}\rho+\LL_{diss}\rho\ .
\end{eqnarray}
We will show that, at the steady state, oscillators at equal distance, on the left and on the right of the central one, are entangled in pairs as depicted in Fig.~\ref{fig0}. 

We remark that the squeezed bath, described by Eq.~\rp{LLsqueez}, is  equal to a standard dissipative bath for a Bogoliubov (squeezed) mode with annihilation operator $c_rb_0-\ee^{\ii\varphi}s_r b_0\da$, that is 
\begin{eqnarray}
\LL_{squeez}\al{\Gamma,\bar n,\bar m}\rho&=&\Gamma\lpg{
(\bar n_r+1)\,\DD\pq{c_rb_0-\ee^{\ii\varphi}s_r b_0\da,c_r b_0\da-\ee^{-\ii\varphi}s_rb_0}
}\nn\\&&\rpg{
+\bar n_r\,\DD\pq{c_r b_0\da-\ee^{-\ii\varphi}s_r\ b_0,c_r b_0-\ee^{\ii\varphi}s_r b_0\da}
}\rho\ ,
\end{eqnarray}
with $\ee^{\ii\varphi}=\bar m/\abs{\bar m}$, $s_r/c_r=\pt{\bar n-\tilde n_r}/\abs{m}$, $c_r^2-s_r^2=1$ and 
\begin{eqnarray}\label{nr}
\tilde n_r&=&\frac{1}{2}\pq{\sqrt{\pt{2\bar n+1}^2-4\abs{\bar m}^2}\ -1}\ .
\end{eqnarray}
Correspondingly, as a result of the symmetric pattern of couplings $\eta_j$ that we have chosen, the central Bogoliubov mode is coupled to the two modes $c_r b_{\pm1}+\ee^{\ii\phi}s_rb_{\mp1}\da$
and in turn these are coupled to similar modes realizing a chain of 
Bogoliubov modes whose annihilation operators are $c_r b_{\pm j}-(-1)^j\ee^{\ii\phi}s_rb_{\mp j}\da$.
Furthermore we note that the Hamiltonian term proportional to the detunings $\Delta_j$, in Eq.~\rp{Hchain} does not induce additional couplings between these Bogoliubov modes. This is a consequence of the chosen antisymmetric structure of the detunings, and of the fact that, in general
the number difference operator $b_j\da b_j-b_k\da  b_k$ is invariant under the effect of a two-mode squeezing transformation.
In conclusion, as discussed in more detail in the next Section, due to the specific symmetries that we have selected, the squeezed bath acts as a standard dissipative reservoir over a chain of Bogoliubov (squeezed) modes.

\section{Steady state entanglement}
\label{steady state}

\begin{figure*}[!t]
\begin{center}
\includegraphics[width=18.cm]{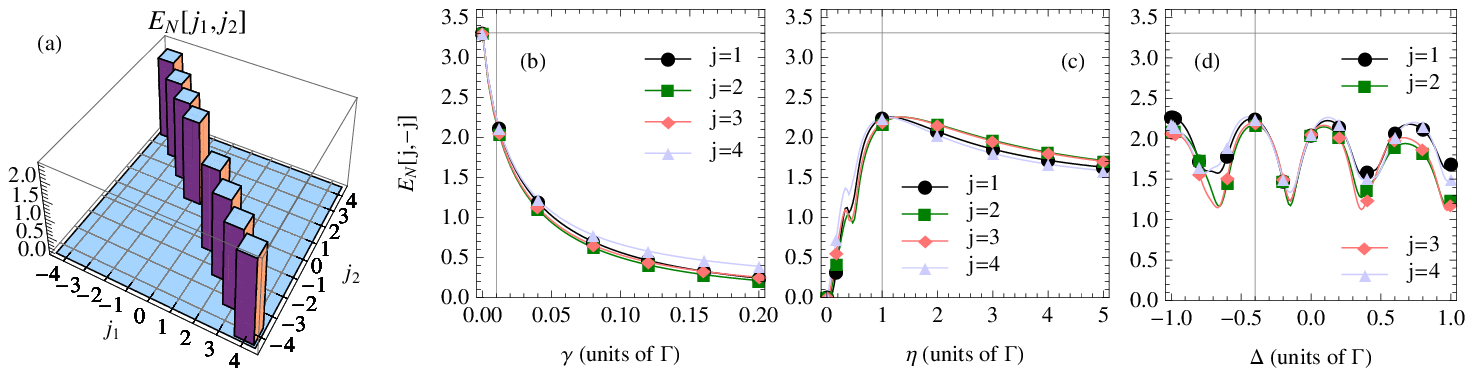}
\end{center}
\caption{(Color online) (a): Steady-state logarithmic negativity, $E_N[j_1,j_2]$, for all the pairs of oscillators in a chain of $9$ oscillators ($N=4$), with couplings and detunings defined in Eq.~\rp{etaDelta}, at the steady state of the dynamics described by Eq.~\rp{Meq0}, with
$\bar n=2$, $\bar m=\sqrt{\bar n(\bar n+1)}$ 
(corresponding to reservoir squeezing of $\sim\!10\,$dB),
 $\eta=\Gamma$, $\Delta=-0.4\,\Gamma$, $\delta=0.1\,\Gamma$, and uniform dissipation with $\gamma_j=0.01\,\Gamma$ and $n_{Tj}=0$ $\forall j$.
(b), (c), and (d): Steady-state logarithmic negativity for the pairs with indices $\{j,-j\}$, as a function of, respectively, the mechanical dissipation $\gamma\equiv\gamma_j$, $\forall j$, the oscillators couplings $\eta$, and of the detuning $\Delta$. The other parameters are as in plot (a). The vertical thin lines in (b), (c) and (d) indicate the parameters corresponding to plot (a). The horizontal thin lines indicate the level of entanglement at $\gamma_j=0$, $\forall j$, as defined in Eq.~\rp{EN0}. 
In all cases the other pairs are not entangled, i.e. $E_N[j_1,j_2]=0$ for $j_1\neq-j_2$.
}\label{figresideal}
\end{figure*}
\begin{figure}[!t]
\begin{center}
\includegraphics[width=8.5cm]{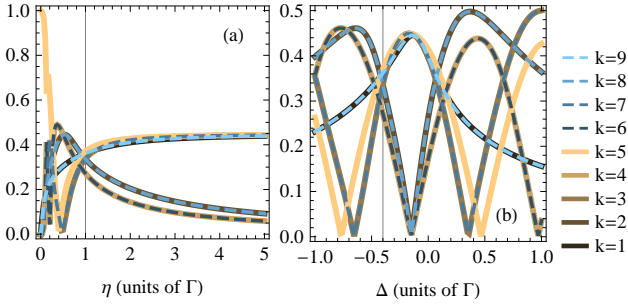}
\end{center}
\caption{(Color online) Projection $\tilde\vv_k^T\cdot\vv_0$ of the eigenmodes of the chain, described by the eigenvectors $\tilde\vv_k$ of the matrix $\MM\Big|_{\Gamma=0}$, defined in Eq~\rp{MM_Gamma0}, over the central oscillator, described by the vector $\vv_0$, defined in Eq.~\rp{vv0}, as a function of (a) the oscillators coupling $\eta$, and (b) the detuning $\Delta$. 
Increasing values of the index $k$ correspond to increasing values of the frequency of the normal modes. The vertical lines indicate the optimal parameters where all the curves converge to a close value, and correspond to the parameters of Fig.~\ref{figresideal} (a) and to the vertical lines in Figs.~\ref{figresideal} (c) and (d). The other parameters are as in Fig.~\ref{figresideal} (a).
}\label{figeigmodes}
\end{figure}
\begin{figure}[!t]
\begin{center}
\includegraphics[width=8.5cm]{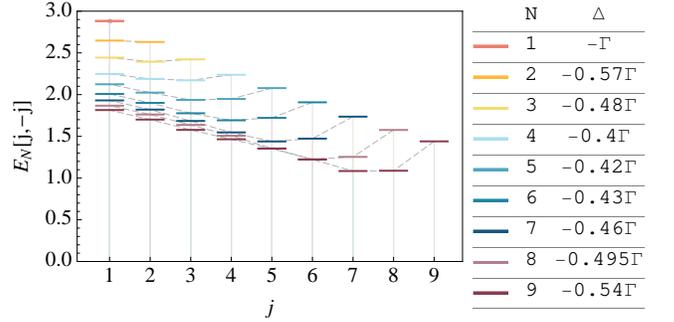}
\end{center}
\caption{(Color online) Steady-state logarithmic negativity, $E_N[j,-j]$, for the pairs of oscillators with indices $j$ and $-j$, for chains of increasing size (from 3 to 19 oscillators, i.e. $N\in\pg{1\cdots 9}$), as detailed in the table. 
For each size the detuning $\Delta$ has been optimized by matching the configuration for which the projections $\tilde\vv_k\cdot\vv_0$ are as close as possible according to the procedure outlined in Fig.~\ref{figeigmodes}. 
For each $j$ the lowest mark corresponds to the entanglement realized with the largest size $(N=9)$. Increasing values of $E_N$ correspond to decreasing sizes.
The dashed lines connect the results obtained with the same value of $N$.
The other parameters are as in Fig.~\ref{figresideal} (a).
}\label{figresN}
\end{figure}

A simple analytical  solution for the steady state is found when $\LL_{diss}=0$ (i.e., $\gamma_j=0$ $\forall j$).
In this case it is useful to study the system in a new representation defined by the unitary operator
\begin{eqnarray}\label{U}
U=\ee^{\frac{r}{2}\pt{\ee^{\ii\varphi}{b_0\da}^2-\ee^{-\ii\varphi}{b_0}^2}}\  \otimes_{j=1}^N \ee^{(-1)^j\, r\,\pt{\ee^{\ii\varphi}b_j\da\ b_{-j}\da-\ee^{-\ii\varphi}b_j\ b_{-j}}}\ ,
\end{eqnarray}
which realizes the Bogoliubov transformation
\begin{eqnarray}
U\ b_0\ U\da&=&\cosh\pt{r}\ b_0+\ee^{\ii\varphi}\sinh\pt{r}\ b_0\da
\nn\\
U\ b_j\ U\da&=&\cosh\pt{r}\ b_j+(-1)^j\ee^{\ii\varphi}\sinh\pt{r}\ b\da_{-j} \  ,
\end{eqnarray}
with $\tanh(r)=\pt{\bar n-\tilde n_r}/{\abs{\bar m}}$ and $\tilde n_r$ defined in Eq.~\rp{nr}.
We note that $\tilde n_r=0$ when $\abs{\bar m}=\sqrt{\bar n(\bar n+1)}$, that is, when the reservoir is in a pure squeezed state.

In the new representation the system Hamiltonian remains unchanged as a result of the symmetry that we have chosen. 
The master equation for the transformed density matrix $\tilde\rho=U\da\rho\ U$ is, therefore, given by 
\begin{eqnarray}\label{Meq01}
\dot{\tilde\rho}&=&-\frac{\ii}{\hbar}\pq{H_{chain},\tilde \rho}
\nn\\&&
+\Gamma\pg{
(\tilde n_r+1)\DD\pq{b_0,b_0\da}+\tilde n_r\DD\pq{b_0\da,b_0}
}\tilde\rho
\end{eqnarray}
that has the same structure of Eq.~\rp{Meq0} but with $\bar m=0$, and $\tilde n_r$ in place of $\bar n$.
This master equation describes a system interacting with a thermal environment with $\tilde n_r$ thermal excitations. In this representation, at large times, the system thermalizes and all the oscillators approach the same thermal state with $\tilde n_r$ excitations
$\tilde\rho_{st,j}\al{r}=\sum_n\frac{1}{1+\tilde n_r}\pt{\frac{\tilde n_r}{1+\tilde n_r}}^n\kb{n}{n}$
(equal for all $j$), so that the global steady state is given by
\begin{eqnarray}\label{tildest}
\tilde\rho_{st}=\otimes_{j=-N}^N \,\tilde\rho_{st,j}\al{r}\ ,
\end{eqnarray}
which is equal to the vacuum when $\tilde n_r=0$, that is when the state of the driving reservoir is pure.
We have to remark, however, that although this thermal state is always a steady solution, in certain situations it is not unique. If for example the matrix of coefficients, $\MM$, of the  system of equations, $\dot\vbe(t)=\MM\,\vbe(t)$, for the evolution of the average field operators $\vbe=\pt{\av{b_{-N}},\cdots,\av{b_0},\cdots,\av{b_N}}^T$, corresponding to Eq.~\rp{Meq01}, has eigenvalues with zero real part, then the corresponding eigenmodes do not dissipate, and the steady-state solution is not unique. This happens, for example, when there exists an eigenvector, $\tilde \vv_k$, of the matrix 
\begin{eqnarray}\label{MM_Gamma0}
\MM\Big|_{\Gamma=0}=-\ii\pt{\mat{cccccc}
{
-\Delta_{N}&\eta_{N} & 0&  \cdots & & 0 \\
\eta_{N}     & \ddots & & & &\vdots\\
0             & &   \eta_1 & 0 &\eta_{1} & 0\\  
 \vdots    & &                &                &\ddots  &   \eta_N \\
 0           &  &       \cdots&   0 &    \eta_N        & \Delta_{N}
}}\ ,
\end{eqnarray} 
that is orthogonal to the vector 
\begin{eqnarray}\label{vv0}
\vv_0=\pt{0,\cdots1,\cdots 0}^T
\end{eqnarray}
that corresponds to the central oscillator, i.e., $\tilde\vv_k^T\cdot \vv_0=0$. In this case, in fact, the corresponding  eigenmode remains unaffected by the thermal reservoir. A specific realization of this situation is observed, for example, when all the oscillators are resonant $\Delta_j=0$.
If instead all the eigenvectors, $\tilde\vv_k$, of  $\MM\Big|_{\Gamma=0}$ are non-orthogonal to $\vv_0$, i.e., $\tilde\vv_k^T\cdot \vv_0\neq0$, then the steady state identified above is unique. 
We have checked that this condition is fulfilled when, for example, the couplings are uniform and the detunings vary linearly according to the relations 
\begin{eqnarray}\label{etaDelta}
\eta_j&=&\eta\ , \ \ \ \ \ \  \ \ \ \ \ \ {\rm for }\  j\in\pg{1\cdots N}\ ,
\nn\\
\Delta_j&=&\Delta+ j\,\delta\ , \ \ \ \ \ \ {\rm for }\  j\in\pg{1\cdots N}\ . 
\end{eqnarray}
We anticipate that the specific results that we discuss hereafter are obtained for chains which follow this structure.

Before studying the system dynamics under more general conditions we have to analyze the entanglement properties of the steady state that we have just identified, and that is valid when $\LL_{diss}=0$. Specifically, in the original representation, the thermal steady state in Eq.~\rp{tildest}, reduces to $\rho_{st}=U\,\tilde\rho_{st}\,U\da$, with $U$ defined in Eq.~\rp{U}, that corresponds to a factorized state of the form $\rho_{st}=\otimes_{j=0}^N\, \rho_{st,j}$ with  the central oscillator in a squeezed thermal state 
\begin{eqnarray}\label{rhost0}
\rho_{st,0}=U\tilde\rho_{st,0}\al{r}U\da \ ,
\end{eqnarray}
and all the pairs of oscillators with opposite indices, $j$ and $-j$, in two-mode squeezed thermal states 
\begin{eqnarray}\label{rhostj}
\rho_{st,j}=U\,\tilde\rho_{st,j}\al{r}\otimes\tilde\rho_{st,-j}\al{r}\,U\da \ ,
\end{eqnarray}
that is the same for all the $j$ regardless of the  size of the chain. We note that these density matrices describe pure two-mode squeezed states when $\tilde n_r=0$.
The corresponding logarithmic negativity~\cite{Vidal,Plenio05} between each pair of oscillators can be expressed in terms of the corresponding correlation functions (see App.~\ref{correlations})  and is given by 
\begin{eqnarray}\label{EN0}
E_N[j,-j]={\rm max}\pg{0,-\log_2S}\ ,
\end{eqnarray}
where $S$ is defined in Eq.~\rp{S},
which shows that the strength of the pairwise entanglement increases with the degree of squeezing of the driving reservoir.
We remark that this value is independent from the size of the chain and from the parameters $\eta_j$ and $\Delta_j$.

The results that we have discussed so far have been found for $\LL_{diss}=0$. When this condition is not satisfied it is, in general, not possible to identify a simple analytical solution. In this case we resort to the numerical evaluation of the steady state and we characterize its entanglement properties in terms of the logarithmic negativity~\cite{Vidal,Plenio05} for different pairs of oscillators $E_N[j_1,j_2]$. 
In general finite values of $\gamma_j$ tend to reduce the entanglement of the pairs, which, however, remain visible as long as the thermal noise is not too strong, as depicted in Fig.~\ref{figresideal} (a) and (b). In any case, {\em only} oscillators with opposite indices can be entangled, i.e. $E_N[j_1,j_2]=0$ for $j_1\neq-j_2$.
We further note that, when $\gamma_j\neq 0$, the pairwise entanglement depends non-trivially on the structure of the chain. For example, when the chain structure is described by Eq.~\rp{etaDelta}, the entanglement of the pairs is maximized for finite values of  $\eta$, while it oscillates as a function of $\Delta$ (see Fig.~\ref{figresideal} (c) and (d)). In detail, 
the optimal value of the entanglement is observed when the projection $\tilde\vv_k^T\cdot\vv_0$ of the eigenmodes of the chain, $\tilde\vv_k$, over the central oscillator, $\vv_0$, are of similar size as described by Fig.~\ref{figeigmodes}.
In this case, in fact, all the normal modes are equally coupled to the reservoir, and they are all efficiently driven to their steady states.
The behaviour of the steady state as a function of the size of the chain is shown in Fig.~\ref{figresN}. 
As already stated when $\gamma_j=0$ all the pairs share the same degree of entanglement regardless of the size of the chain. In contrast, as shown in Fig.~\ref{figresN}, when $\LL_{diss}\neq 0$, the entanglement of each pair decreases mildly with the number of the oscillators in the chain. We also observe that, similar to the findings of Ref.~\cite{Zippilli13}, the entanglement decreases with the distance of the pair from the central oscillator, and exhibits a weak revival for a few pairs at the ends of the chain.

\begin{figure}[!th]
\begin{center}
\includegraphics[width=8.5cm]{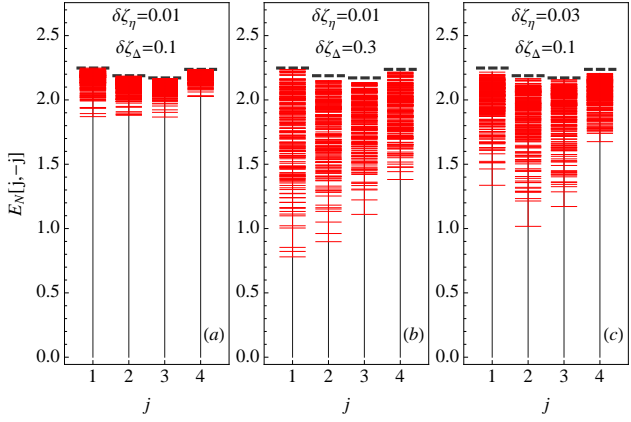}
\end{center}
\caption{
(Color online) Steady-state logarithmic negativity, $E_N[j,-j]$, for the pairs of oscillators with indices $j$ and $-j$. Each plot reports the results for 200 different realizations (red-solid-thin marks) of the dimensionaless random variables $\zeta_{\eta,j}$ and $\zeta_{\Delta,j}$, uniformly distributed in the ranges $\zeta_{\eta,j}\in\pq{-\delta\zeta_\eta,\delta\zeta_\eta}$ and $\zeta_{\Delta,j}\in\pq{-\delta\zeta_\Delta,\delta\zeta_\Delta}$, which describe small fluctuations of the oscillators couplings and detunings about the ideal symmetric situation identified in Sec.~\ref{ideal}, as discussed in the main text. Each plot is evaluated with different values of the range of variability $\delta\zeta_\eta$ and $\delta\zeta_\Delta$ of the random variables, as specified in the figure. 
The black-dashed-thick marks indicate the entanglement in the symmetric configuration, i.e. $\zeta_{\eta,j}=\zeta_{\Delta,j}=0$, $\forall j$.
The other parameters are as in Fig.~\ref{figresideal} (a).
}\label{figresRand}
\end{figure}
We finally discuss the effects of random deviations from the symmetric configuration identified in Sec.~\ref{ideal}.
The outcome of this analysis is reported in Fig.~\ref{figresRand}, where we show the results for the steady state entanglement evaluated considering a Hamiltonian of the form 
\begin{eqnarray}\label{Hchain_rand}
H_{chain}&=&\hbar\sum_{j=1}^N\pt{\Delta+j\delta}\pq{\pt{1+\zeta_{\Delta,j}}b_j\da\ b_j-\pt{1+\zeta_{\Delta,-j}}b_{-j}\da\ b_{-j}}
\nn\\&&
+\hbar\sum_{j=1}^N \eta\lpq{
\pt{1+\zeta_{\eta,j}}\pt{b_{j-1}\ b_j\da+b_{j-1}\da\ b_j}
}\nn\\&&\rpq{
+\pt{1+\zeta_{\eta,-j}}\pt{b_{-j+1}\ b_{-j}\da+b_{-j+1}\da\ b_{-j}}
}\ ,
\end{eqnarray}
where $\zeta_{\eta,j}$ and $\zeta_{\Delta,j}$ are random variables  uniformly distributed in the ranges $\pq{-\delta\zeta_\eta,\delta\zeta_\eta}$ and $\pq{-\delta\zeta_\Delta,\delta\zeta_\Delta}$ respectively. Specifically, Eq.~\rp{Hchain_rand} accounts for deviations from the configuration described by Eq.~\rp{Hchain} with the parameters defined in Eq.~\rp{etaDelta}.
Fig.~\ref{figresRand} demonstrates that the protocol is significantly resilient to small asymmetries. We also note that the protocol is much less sensitive to deviations in the detunings than in the oscillators couplings.

\section{Physical implementation: Reservoir engineering with bichromatic driving fields}
\label{physcalSystems}

Here we analyze how to implement the dynamics discussed in the previous section with actual physical systems. 
We first note that, in principle, one could realize the model described in Sec.~\ref{ideal} by, for example, driving arrays of optical or microwave resonators with the output field of a degenerate parametric oscillator below threshold, which would constitute the squeezed reservoir similar to the proposal discussed in Ref.~\cite{Zippilli14}.
A promising alternative approach, that we explore here, is instead that of realizing a squeezed bath by reservoir engineering using bichromatic driving fields.
The idea of using bichromatic fields to engineer quantum dissipative dynamics has been introduced in the context of trapped ions in Ref.~\cite{Cirac93}. It has been then extended to a number of different physical setups including mechanical resonators~\cite{Rabl04,Mari09,Kronwald13,Kronwald14} and circuit-QED~\cite{Porras12}. All these systems are in principle suitable candidates for the engineering of the dynamics that is the subject of the present proposal. Hereafter we briefly review and extend these techniques based on two-frequency drives to the simulations of our protocols with 
two specific setups. First, we consider an array of mechanical resonators and then one of superconducting microwave resonators. In both cases we identify the parameters for which the dynamics discussed in Sec.~\ref{ideal} can be efficiently simulated.
We characterize the system dynamics in terms of the logarithmic negativity for pairs of oscillators evaluated at the steady state of the system dynamics showing that the results identified in the previous section can be reproduced under specific parameter regimes.

\subsection{Optomechanics: Simulation with an array of mechanical resonators}
\label{S:MRchain}

\begin{figure}[!t]
\begin{center}
\includegraphics[width=8.5cm]{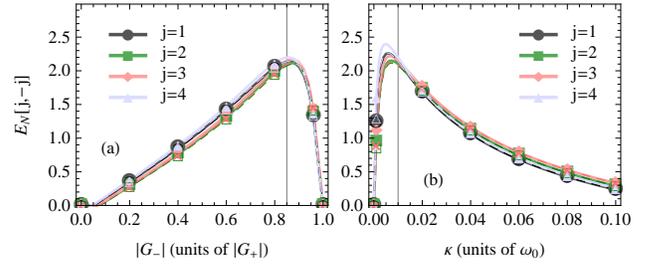}
\end{center}
\caption{(Color online) 
Steady-state logarithmic negativity $E_N[j,-j]$ for the pairs of resonators with opposite indices, in a chain of 9 ($N=4$) mechanical resonators, as a function of (a) $\abs{G_-}$, with $\kappa=0.01\omega_0$, and (b) $\kappa$, with $\abs{G_-}=0.85\,\abs{G_+}$ ($\EE_-=136\omega_0$).
The solid and dashed lines, that are almost indistinguishable because very close to each other, are evaluated from the steady state of, respectively, Eq.~\rp{Meq1} and Eq.~\rp{Meq0} with the parameters defined in Eq.~\rp{effpar1}.
The vertical line in (a) indicates the parameters used in (b) and similarly the one in (b) indicates the value of $\kappa$ used in (a). 
The other parameters are 
$\omega_0=2\pi\times1$GHz, $g=5\times10^{-5}\omega_0$, $\delta=0.2\times10^{-3}\omega_0$,
$\lambda_{cav}=1550$nm, $\abs{G_+}=8\times10^{-3}\omega_0$ ($\EE_+=160\omega_0$), temperature $T=50$mK and $\gamma_j=10^{-5}\omega_0$ $\forall j$. Moreover the chain follows the structure defined by Eq.~\rp{etaDelta} with $\eta=2\times10^{-3}\omega_0$ and $\Delta=0.8\times10^{-3}\omega_0$ that are consistent with the optimal values identified in Fig.~\ref{figeigmodes}. All the other pairs are not entangled, i.e. $E_N[j_1,j_2]=0$ for $j_1\neq-j_2$.
}\label{figresOM1}
\end{figure}
\begin{figure}[!t]
\begin{center}
\includegraphics[width=8.5cm]{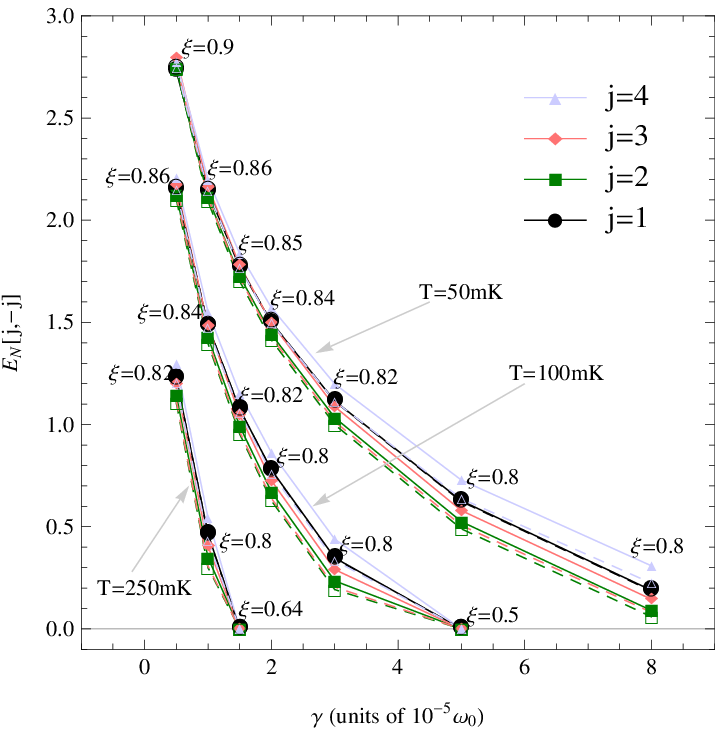}
\end{center}
\caption{(Color online) 
Steady-state logarithmic negativity $E_N[j,-j]$ for pairs of resonators with opposite indices, in a chain of 9 mechanical  resonators as a function of the dissipation rate of the resonators (assumed uniform through the chain, $\gamma_j\equiv\gamma$ $\forall j$), for three different values of the temperature $T$ as specified by the arrows. For each set of points the linearized coupling $G_-$ has been chosen in order to optimize the entanglement of the pairs, and the corresponding  values of $\xi=\abs{G_-}/\abs{G_+}$ are indicated in the plot. 
The solid and dashed lines connect the results found from the steady state of, respectively, Eq.~\rp{Meq1} and Eq.~\rp{Meq0}, where in the latter the bath parameters are defined in Eq.~\rp{effpar1}. The other parameters are as in Fig.~\ref{figresOM1}. All the other pairs are not entangled, i.e. $E_N[j_1,j_2]=0$ for $j_1\neq-j_2$.
}\label{figresOM2}
\end{figure}

In this section we describe how to engineer the dynamics introduced in Sec.~\ref{steady state} with a chain of mechanical resonators similar to those discussed in~\cite{Habraken12,Schmidt12,Ludwig13}, which are based on optomechanical crystal implementations~\cite{Eichenfield,Safavi-Naeini10,Safavi-Naeini11,Chan,Gavartin,Safavi-Naeini14}, where mechanical and optical modes are defined by localized defect in planar artificial quasi-periodic nanostructured crystals. The system hence consists of an array of localized mechanical modes in a phononic-crystal. Phononic excitations can hop between nearby mechanical modes~\cite{Ludwig13} realizing the chain.
The squeezed reservoir is, instead, realized by having the central mechanical mode coupled to an optical mode~\cite{Gavartin,Safavi-Naeini14} that, in turn, is driven by a two-tone field realizing a reservoir-engineering scheme similar to the one investigated in~\cite{Mari09,Kronwald13}. 
The corresponding linearized optomechanical dynamics can be described by a master equation for the system density matrix $\rho$ of the form (see App.~\ref{MRchain} for a detailed derivation)
\begin{eqnarray}\label{Meq1}
\dot{\rho}=-\frac{\ii}{\hbar}\pq{H,\rho}+\pt{\LL_{a}+\LL_{diss}}\rho\ ,
\end{eqnarray}
where, the cavity dissipation at rate $\kappa$ is described by $\LL_{a}=\kappa\DD\pq{a,a\da}$, with $\DD$ defined in Eq.~\rp{DD}, $\LL_{diss}$ is introduced in Eq.~\rp{LLdiss}, and the Hamiltonian is given by 
\begin{eqnarray}\label{H}
H=H_{chain}+\hbar\pq{a\da\pt{G_+ b_0+G_- b_0\da }
+a\pt{G_+^*b_0\da+G_-^*b_0 }}\ .
\end{eqnarray}
Here $H_{chain}$ is defined as in Eq.~\rp{Hchain}, the operators $a$ and $b_j$ account for the fluctuations of, respectively, the optical cavity and the mechanical resonators about their corresponding average fields, and the linearized coupling strengths $G_+$ and $G_-$ are proportional to the intensity of the driving fields on the red and blue sideband respectively (their specific dependence on the other system parameters is discussed in App.~\ref{linearization}). 

As discussed in~\cite{Kronwald13}, when $\abs{G_+}>\abs{G_-}$ (which guarantees the stability of this model, keeping however in mind that in certain cases, when $\abs{G_+}\sim\abs{G_-}$,  non resonant processes that have been neglected in this effective model (see App.~\ref{MRchain}) can modify the system stability~\cite{Jie}), this model describes the cooling of the mechanical Bogoliubov mode of the central oscillator given by $\hat B_0=\pt{G_+ b_0+G_- b_0\da}/{\sqrt{\abs{G_+}^2-\abs{G_-}^2}} $. If the cooling dynamics is sufficiently fast to overcome the thermal noise then such a mode can be cooled to its ground state, that corresponds to a squeezed state for the original mode. In turn the other oscillators of the chain are driven to an entangled steady state following a dynamics similar to the one discussed in Sec.~\ref{steady state}.
In particular, if the decay rate of the cavity field, $\kappa$, is sufficiently large, the degrees of freedom of the optical cavity can be adiabatically eliminated and the system can be described by a model equal to Eq.~\rp{Meq0} with
\begin{eqnarray}\label{effpar1}
\Gamma&=&\frac{\abs{G_+}^2-\abs{G_-}^2}{\kappa}
\nn\\
\bar n&=&\frac{\abs{G_-}^2}{\abs{G_+}^2-\abs{G_-}^2}
\nn\\
\bar m&=&-\frac{G_+^*\,G_-}{\abs{G_+}^2-\abs{G_-}^2}\ .
\end{eqnarray}

Here we consider systems similar to the ones investigated in~\cite{Eichenfield,Chan,Safavi-Naeini10,Safavi-Naeini11,Gavartin,Safavi-Naeini14}, where the mechanical modes are at gigahertz frequencies while the optical ones are in  the near infrared. In detail we choose $\omega_0=2\pi\times1$GHz and the wavelength for the cavity field $\lambda_L=1550$nm. 
Figs.~\ref{figresOM1} and \ref{figresOM2}  show that the model of Sec.~\ref{ideal} describes with high accuracy the dynamics of this system in the corresponding parameter regime.
We observe, in fact, that dashed and solid lines, corresponding to the results obtained  from the steady state of, respectively, Eqs.~\rp{Meq0} and \rp{Meq1},  are always very close.
The pairwise entangled steady-state is achieved whenever the effective squeezed reservoir is efficiently realized. The parameter regimes for which the bichromatic drives squeeze  efficiently  a mechanical oscillator have been discussed in detail in Ref.~\cite{Kronwald13}.
In particular, the squeezing is optimized at specific values of $\abs{G_-}/\abs{G_+}$, and of the decay rate of the cavity field. Correspondingly we observe strong pairwise entanglement as described by Figs.\ref{figresOM1} (a) and (b). In turn,
when the optimal conditions for $G_\pm$ and $\kappa$ are satisfied, the entanglement is achieved if the chain follows the symmetries identified in Sec.~\ref{ideal}, and the couplings and detunings are optimized as discussed in the previous section. In this case, all the pairs of resonators with indices $j$ and $-j$ exhibit roughly the same value of the logarithmic negativity $E_N[j,-j]$, and in general larger entanglement is achieved for larger quality factor $\omega_0/\gamma_j$ of the resonators and for smaller temperatures as shown in Fig.~\ref{figresOM2}.

\subsection{Circuit QED: Simulation with an array of superconducting coplanar waveguide resonators}
\label{S:SRchain}

\begin{figure}[!t]
\begin{center}
\includegraphics[width=8.5cm]{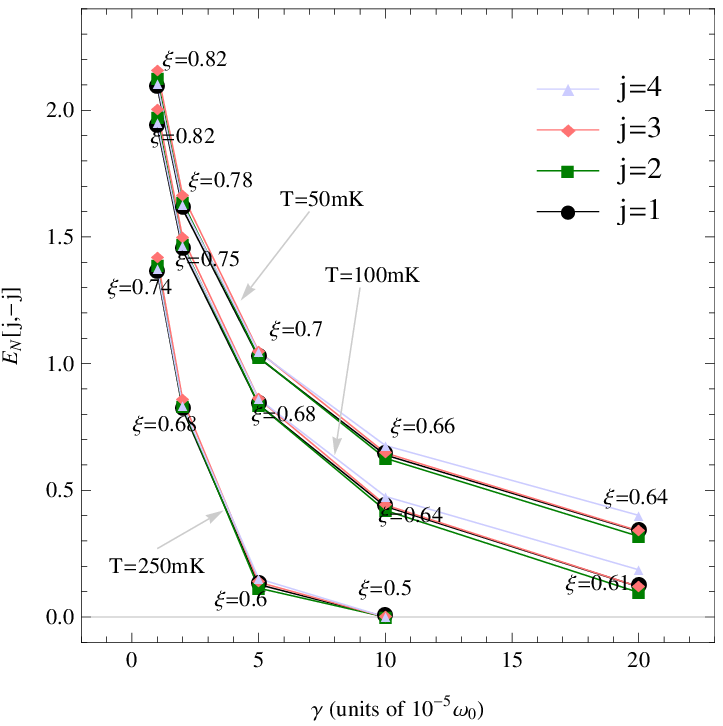}
\end{center}
\caption{(Color online) 
Steady-state logarithmic negativity $E_N[j,-j]$ for pairs of resonators with opposite indices, in a chain of 9 coplanar waveguide resonators as a function of the dissipation rate of the resonators (assumed uniform through the chain, $\gamma_j\equiv\gamma$ $\forall j$), for three different values of the temperature $T$ as specified by the arrows. The coupling strength $G_-$ has been chosen in order to optimize the corresponding entanglement for each set of points, and the corresponding  values of $\xi=\abs{G_-}/\abs{G_+}$ are indicated in the plot. 
The results are evaluated from the steady state of Eq.~\rp{Meq0} with the bath parameters defined in Eq.~\rp{effpar1}, and with the effective coupling $G_\pm$ defined in Eq.~\rp{GpmS}.
The other parameters are
$\omega_0=2\pi\times5$GHz, $g=0.06\omega_0$, $\eta=2\times10^{-3}\omega_0$, $\Delta=0.8\times10^{-3}\omega_0$, $\delta=0.2\times10^{-3}\omega_0$, $\epsilon=2\omega_0$, $\abs{G_+}=6\times10^{-3}\omega_0$ ($\EE_+=0.25\omega_0$)
$\kappa=0.02\omega_0$. All the other pairs are not entangled, i.e. $E_N[j_1,j_2]=0$ for $j_1\neq-j_2$.
}\label{figresSR2}
\end{figure}

A different system that can be used to engineer the dynamics described in Secs.~\ref{ideal} and \ref{steady state} consists of a chain of capacitively coupled microwave coplanar waveguide (CPW) resonators~\cite{Houck,Schmidt13}. This system is attractive because CPW resonators can be efficiently coupled to superconducting qubits, hence realizing a dynamics similar to the one discussed in Ref.~\cite{Zippilli13}, where also the qubits get entangled in the steady state.

In this case, the squeezed reservoir can be engineered by coupling only the central resonator to a superconducting qubit. Specifically one can implement the scheme described in~Ref.~\cite{Porras12}, where a superconducting flux qubit~\cite{Mooij,Orlando} with tunable gap~\cite{Paauw,Fedorov}, is coupled to a CPW resonator at frequency $\omega_0$, and the gap is modulated by a bichromatic field. We remark that in principle similar dynamics can be realized with different kinds of superconducting qubits.
At the degeneracy point the flux qubit is described (following the notation in Ref.~\cite{Porras12}) as an effective spin 1/2 in a modulated magnetic field
\begin{eqnarray}
H_q=\hbar\pq{\frac{\epsilon}{2}-\EE_+\cos(\epsilon+\omega_0)-\EE_-\cos(\epsilon-\omega_0)}\sigma_z\ ,
\end{eqnarray}
while the interaction with the resonator, the creation and annihilation operators of which are $b_0$ and $b_0\da$, is given by
\begin{eqnarray}
H_I=\hbar\,g\,\pt{\sigma_++\sigma_-}\pt{b_0+b_0\da}\ ,
\end{eqnarray}
where $\sigma_\pm$ are the lowering and rising operators for the effective spin.  
Here we assume that this resonator is the central element of a chain of CPW resonators described by a Hamiltonian equal to Eq.~\rp{barHchain1}.
In the interaction picture with respect to $H_q+\hbar\,\omega_0\sum_j\, b_j\da b_j $ one finds that, similar to the derivation in Ref.~\cite{Porras12}, the total system Hamiltonian 
at lowest order in $\EE_\pm/(\epsilon\pm\omega_0)$, and retaining only the dominant resonant terms, under the condition $\kappa,\ g\,\EE_\pm/(\epsilon\pm\omega_0)\ll\omega_0$, reduces to 
\begin{eqnarray}\label{Hs}
H_s=H_{chain}+\hbar\,\sigma_+\pq{G_+ b_0+G_- b_0\da }+h.c.
\end{eqnarray}
where 
\begin{eqnarray}\label{GpmS}
G_\pm=\frac{g\,\EE_\pm}{\epsilon\pm\omega_0} \ ,
\end{eqnarray}
and $H_{chain}$ is defined as in Eq.~\rp{Hchain}.
Also in this case, if the qubit decays fast enough, then the Bogoliubov mode $\hat B_0=\pt{G_+ b_0+G_- b_0\da}/{\sqrt{\abs{G_+}^2-\abs{G_-}^2}}$  can be efficiently cooled.
Fast relaxation, that should overcome the natural dephasing of the device, can be engineered by constructing the sample with a nearby transmission line which serves as an additional dissipation channel. If the resulting relaxation rate, that we denote as $\kappa$, is sufficiently large then the qubit remains mostly in its ground state. In this case the effective spin can be approximated as a harmonic oscillator. Thereby the system description reduces to a model equal to Eq.~\rp{Meq1}, where the bosonic operators $a$ and $a\da$ have replaced $\sigma_-$ and $\sigma_+$, respectively. 
We have verified that using this model, with the parameters that we will discuss below, the population of the bosonized qubit is always very small ($\lesssim0.02$), hence providing a justification of the results presented hereafter. Specifically, under these conditions, the degrees of freedom of the qubit can be adiabatically eliminated, and also in this case the system is described by a model equal to Eq.~\rp{Meq0}, with the bath parameters defined as in Eq.~\rp{effpar1}. 
 
In detail, we consider system parameters consistent with state-of-the-art circuit-QED experiments~\cite{Schmidt13}.
We focus on  resonators with frequencies of $\omega_0=2\pi\times 5$GHz, and quality factors $Q\sim10^5$. The frequency of the central qubit is $2\pi\times 10$GHz and its coupling to the resonator is $g\sim2\pi\times300$MHz. Moreover we employ a large decay rate of the qubit $\kappa\sim 2\pi\times100$MHz, much larger than its natural dephasing rate (the largest dephasing time in flux qubits is of $\sim10\mu$s~\cite{Stern}, corresponding to rates of tens of kilohertz) which here is therefore neglected.
Assuming a chain of resonators described by coupling strengths and detunings consistent with those identified in Sec.~\ref{steady state}, the steady-state results are very similar to those found for the mechanical resonators as depicted in Fig.~\ref{figresSR2}.

\section{Discussion and Conclusions} 
\label{conclusions}

We have shown that a harmonic chain with nearest-neighbour coupling can be driven to a non-trivial steady state exhibiting a series of nested entangled pairs covering the whole chain when only the central oscillator is coupled to a single mode squeezed bath.
Ideally, when the effects of additional noise are negligible ($\gamma_j\sim0$ $\forall j$), the entanglement of all the pairs is the same regardless of the size of the chain, and its strength is directly related to the degree of squeezing of the reservoir (see Eq.~\rp{EN0}). 
We have shown that the efficiency for the production of the pairwise entanglement is strongly dependent on the structure of the chain. In particular we have identified the symmetry conditions which are necessary for the observation of this phenomenon, and we have described, in a simple specific realization, how to optimize the entanglement based on the study of the normal modes of the chain.
Finally, we have studied how to simulate this effect, using bichromatic drives, and with actual physical systems, hence demonstrating that optomechanical as well as circuit-QED systems offer the potentiality to harness the discussed dynamics in parameter regimes that are within the reach of near-future experiments.

\subsection{Alternative implementations}

Let us now comment on the possibility to simulate the dynamics described in this article
with other physical systems. 
We remark that any system that can be mapped, in some limit to a harmonic chain is a suitable candidate. 
Furthermore, we observe that  in principle the squeezed reservoir can be realized using the output field of a degenerate parametric oscillator operating below threshold (similar dynamics have been studied for example in Ref.~\cite{Zippilli14}). It consists of a continuous-wave squeezed field mode that, hence, exhibits entanglement between the spectral components at opposite sideband frequencies~\cite{Zippilli15}. In this configuration, the entangled pairs production that we have analyzed can be seen as a means to extract and spatially separate the entanglement contained in a single mode squeezed continuous field. 

Other systems which can be manipulated with bichromatic drives, and that can be exploited in our protocol are listed below. 

\paragraph{Array of optical cavities with optomechanical manipulation:}

First we note that in principle one can envisage an implementation, complementary to the one studied in Sec.~\ref{S:MRchain} which makes use of an array of optical cavities where only the central one interacts with a mechanical resonator. The central cavity is driven by a two-tone field to engineer the squeezed bath similar to the results discussed in Ref.~\cite{Kronwald14}. The system is thus described, with minor modifications, by the model of Sec.~\ref{S:MRchain} (see also App.~\ref{OCchain}).
In this case, the chain of optical cavities can also be realized with arrays of microcavities which are coupled via evanescent waves~\cite{Zhang12} or by optical-fibers~\cite{Derntl}.

\paragraph{Arrays of mechanical resonators:}

Apart from optomechanical crystals, other realization of mechanical arrays based on nano- and micro-mechanical  cantilevers, beams, and membranes, and that can be manipulated with either optical, electrical, or magnetic forces, could be used for our purpose~\cite{Fu,Rabl,Okamoto,Eichler,Matheny,Biswas}. 
In particular electromechanical arrays are very promising systems that are actively investigated in many experiments. 
They can be efficiently coupled to superconducting microwave cavities~\cite{Rocheleau,Palomaki} which can, thus, be used to realize a model formally equal to the one described in Sec.~\ref{S:MRchain}. Moreover squeezed reservoirs for an electromechanical resonator have been recently realized using two-frequency drives~\cite{Wollman,Pirkkalainen}.

\paragraph{Array of superconducting microwave resonators with Electro-optical manipulation:}

Another intriguing possible implementation relies on the realization of 
the squeezed reservoir, for a chain of CPW resonators, by coupling the central oscillator to an optical cavity via an electro-optic modulator~\cite{Tsang}. It consists of a second-order nonlinear-optical medium which is simultaneously placed inside an optical cavity and between the plate of a capacitor which is connected to the CPW resonator. It induces a voltage-dependent phase shift to the cavity optical field, whose corresponding interaction Hamiltonian is, thus, equal to the standard optomechanical one. Consequently,
if the cavity is driven by a two-tone field, this system is described by a model equal to the one discussed in Sec.~\ref{S:MRchain} and in App.~\ref{MRchain}. The experimental feasibility of similar devices is still to be demonstrated, and the actual implementation of these ideas would, most probably, require careful studies of improved system design in order to achieve sufficiently large coupling strengths. 

\paragraph{Trapped ions:}

Trapped ions are highly controllable systems that have been proved to be an efficient platform for quantum simulations and engineering~\cite{Blatt}, and for which the squeezed reservoir, engineered with two-frequency drives, has been already experimentally realized~\cite{Kienzler}. In this case, however, the challenging aspect is probably the design of the chain. In fact, the long range nature of the Coulomb interaction between the ions makes the realization of chains with only nearest neighbour coupling problematic.
In any case, it is worth observing that, this aspect does not necessarily prevent the possibility of
applying our protocol to trapped ions. Indeed, an interesting question, that deserves further investigation, is how the dynamics that we have studied is affected when long range interactions are taken into account.

\subsection{Outlook}

In conclusion we observe that, while we have shown how to optimize the entanglement of the pairs under the specific simple conditions defined by Eq.~\rp{etaDelta}, 
a still open and compelling question is the identification of the optimal structure of the chain (coupling strengths and frequencies of the oscillators) which could result in the maximum steady-state entanglement depending on the specific statistical properties of the driving reservoir.
Moreover, we note that it would be interesting, in relation to the analysis of Sec.~\ref{S:SRchain}, to extend the presented scheme to the situation in which each resonator interacts with a superconducting qubit, such that also pairs of qubits could get entangled at the steady state, analogously to the findings discussed in~\cite{Zippilli13}. Similar considerations in the context of trapped ions could, instead, lead to the creation of entanglement between the internal states of the ions. 
Finally, another very challenging, but fascinating, future research direction would be the investigation of possible extensions of this protocol for the generation of more complex quantum states, in higher dimensional arrays and with arbitrary-range interactions which could be useful, for example, in the context of one-way quantum computation.

\section*{Acknowledgments}

This work has been supported by the European Commission (ITN--Marie Curie project cQOM and FET--Open Project iQUOEMS) and by MIUR (PRIN 2011).

\appendix

\section{Steady state correlations when $\LL_{diss}=0$}\label{correlations}

Here we study the correlations between the operators of the oscillators at the steady state of the dynamics described by Eq.~\rp{Meq0} when  $\LL_{diss}=0$.
Since the system is Gaussian, the following expressions fully characterize the steady state.

Specifically, the correlation matrix for the central oscillator, whose elements are written in terms of the vector of operators $\vb_0=\pt{b_0,b_0\da}$ as $\pg{C_0}_{\ell,\ell'}={\rm Tr}\pq{\pg{\vb_0}_\ell\, \pg{\vb_0}_{\ell'}\,\rho_{st,0} }$, where $\rho_{st,0}$ is defined in Eq.~\rp{rhost0} and $\ell,\ell'$ are vector indices not to be confused with the indices of the oscillators, is 
\begin{eqnarray}\label{A0}
C_0=\pt{
\begin{array}{cc}
m  &  n+1 \\
n  & m^* 
\end{array}
}\ .
\end{eqnarray}
This corresponds to a squeezed state if $\abs{m}>n$, that is when the reservoir is actually squeezed. The correlation matrix for two modes $j$ and $-j$, the elements of which are written in terms of the vector of operators $\vb_j=\pt{b_j,b_{-j},b_j\da,b_{-j}\da}$ as
$\pg{C_j}_{\ell,\ell'}={\rm Tr}\pq{\pg{\vb_j}_\ell\, \pg{\vb_j}_{\ell'}\, \rho_{st,j}}$, with $\rho_{st,j}$ defined in Eq.~\rp{rhostj}, is, instead given by
\begin{eqnarray}\label{Aj}
C_j=\pt{
\begin{array}{cccc}
0 & (-1)^j m  &  n+1 &0\\
(-1)^j m& 0&0&n+1\\
n&0&0&(-1)^j m^*\\
0&n  & (-1)^j m^*&0
\end{array}
}\end{eqnarray}
that corresponds to an entangled state when $\abs{m}>n$, and describes a pure state when $\abs{m}=\sqrt{n(n+1)}$. 
All the other correlation functions between different oscillators are zero.

\section{Optomechanical chain driven by a bichromatic field}
\label{MRchain}

We consider $2N+1$ resonators, in a configuration similar to the one discussed in section~\ref{ideal}, and described by a Hamiltonian of the form 
\begin{eqnarray}\label{barHchain1}
\overline H_{chain}&=&\hbar\,\omega_0\ b_0\da b_0+\hbar\sum_{\xi=\pm1}\, \sum_{j=1}^N\ \pt{\omega_0+\xi\Delta_j}\ b_{\xi j}\da b_{\xi j}
\nn\\&&
+\hbar\sum_{\xi=\pm1}\, \sum_{j=1}^N\ \eta_j
\pq{b\da_{\xi\pt{j-1}}+b_{\xi\pt{j-1}}}
\pq{b_{\xi j} + b\da_{\xi j}   }
\end{eqnarray}
where positive and negative indices indicate resonators that are placed, respectively, to the left and to the right about the central one whose index is $j=0$. The central oscillator is coupled to an optical cavity at frequency $\omega_c$ with annihilation and creation operators $a$ and $a\da$,  via an optomechanical interaction, with strength $g$, described by 
\begin{eqnarray}\label{Hom1}
H_{om}=\hbar\,g\,a\da a\pt{b_0\da+b_0}\ .
\end{eqnarray}
The cavity is driven by two laser fields at the mechanical sideband frequencies $\omega_L\pm\omega_0$, where the central frequency is shifted from the cavity resonance by a small quantity $$\delta_c=\omega_c-\omega_L\ ,$$ 
the specific value of which, as specified below, is chosen to be opposite to the shift of the cavity resonance due to the optomechanical interaction.
The corresponding Hamiltonian term describing the cavity driving, in a reference frame rotating at the frequency $\omega_L$, is given by
\begin{eqnarray}
H_{d}&=&\hbar\,a\da\pq{\EE_-\, \ee^{-\ii\,\omega_0\, t} +\EE_+\, \ee^{\ii\,\omega_0\, t}} +h.c. \ ,
\end{eqnarray}
where $h.c.$ indicates the Hermitian conjugates.
We further take into account the dissipation of the cavity field at rate $\kappa$, which is
described by the standard Lindblad operator $\LL_{a}=\kappa\DD\pq{a,a\da}$, where $\DD$ is defined in Eq.~\rp{DD}, and the thermal noise at rate $\gamma_j$, due to thermal reservoirs with $n_{Tj}$ thermal excitations, that is described by Eq.~\rp{LLdiss}. 

The total system dynamics is therefore described by the master equation
\begin{eqnarray}
\dot\rho&=&-\frac{\ii}{\hbar}\pq{\hbar\,\delta_c\, a\da a +\overline H_{chain}+H_{om}+H_d,\rho}
+\pt{\LL_{a}+\LL_{diss}}\rho\ .
\nn\\
\end{eqnarray}

\subsection{Linearized dynamics, approximation in powers of $g$ and elimination of the fast rotating terms}
\label{linearization}

The optomechanical non-linear dynamics can be linearized around the steady state average amplitudes of the cavity field, $\alpha(t)$, and of the mechanical resonators, $\beta_j(t)$, which fulfill the relations
\begin{eqnarray}
\dot\alpha(t)&=&-\pt{\ii\,\delta_c+\kappa}\alpha(t)-2\,\ii\,g\,\alpha(t){\rm Re}\pg{\beta_0(t)}
\nn\\&&-\ii\pq{\EE_-\ee^{-\ii\,\omega_0\, t}+\EE_+\ee^{\ii\,\omega_0\, t}}
\nn\\
\dot\vbe(t)&=&-\pt{\WW+\ii\,\omega_0\,\id}\vbe(t)-\ii\,g\abs{\alpha(t)}\vv_0\ ,
\end{eqnarray}
where we have introduced the vectors 
\begin{eqnarray}\label{vbvv0}
\vbe(t)&=&\pt{\beta_{-N}(t),\cdots\beta_0(t),\cdots\beta_N(t)}^T \ ,
\nn\\
\vv_0&=&\pt{0,\cdots 1,\cdots 0}^T \ , 
\end{eqnarray}
and the matrix
\begin{eqnarray}\label{WW}
\WW=\pt{\mat{cccccc}
{
w_{-N}&\ii\,\eta_{N} & 0&  \cdots & & 0 \\
\ii\,\eta_{N}     & \ddots & & & &\vdots\\
0             & &    \ii \,\eta_1 & w_{0} &\ii\,\eta_{1} & 0\\  
 \vdots    & &                &                &\ddots  &  \ii\, \eta_N \\
 0           &  &       \cdots&   0 &    \ii\,\eta_N        & w_{N}
}}\ ,
\end{eqnarray} 
with $w_{\pm j}=\pm\ii\Delta_j+\gamma_{\pm j}$ for $j\in\pg{0,1,\cdots N}$.
In the shifted representation described by the density matrix $\rho'=U(t)\da \,\rho\, U(t) $, where
$U(t)$ is the displacement operator $U(t) =D_{\alpha(t)}\otimes_{j=-N}^ND_{\beta_j(t)}$,
with
$D_{\alpha(t)}=\ee^{{\alpha(t)}\,a\da-{\alpha(t)}^*\,a}$
and
$D_{{\beta_j(t)}}=\ee^{{\beta_j(t)}\,b_j\da-{\beta_j(t)}^*\,b_j}$,
that realizes the transformations
\begin{eqnarray}
U(t) \da\ a\ U(t) &=&a+{\alpha(t)}
\nn\\
U(t) \da\ b_j\ U(t) &=&b_j+{\beta_j(t)}\ ,
\end{eqnarray}
the optomechanical interaction Hamiltonian can be approximated as
\begin{eqnarray}
H'_{om}&=&U(t)\da\, H_{om} \,U(t)\simeq   2\,\hbar\,g\,a\da a\, {\rm Re}\pg{\beta_0(t)}
\nn\\&&+\hbar\,g\pq{\alpha(t)\,a\da+\alpha(t)^*\,a}\pt{b_0\da+b_0} \ .
\end{eqnarray}
The first term is a time dependent shift of the optical frequency, while the second one accounts for the linearized optomechanical interaction.
We note that the zero-frequency component of the coefficients $\beta_0(t)$, that we indicate with the symbol $\beta_0^{DC}$, and that contributes to a constant shift of the cavity resonance, can be taken into account by defining a renormalized cavity frequency 
\begin{eqnarray}
\omega_c'=\omega_c+2\,g\,{\rm Re}\pg{\beta_0^{DC}}\ .
 \end{eqnarray}
In particular we set the central frequency of the driving field $\omega_L$ to be resonant with this shifted cavity frequency
\begin{eqnarray}
\omega_L=\omega_c' \ .
\end{eqnarray}
Correspondingly we define the time-dependent part of $\beta_0(t)$ as
\begin{eqnarray}
\bar\beta_0(t)=\beta_0(t)-\beta_0^{DC}\ .
\end{eqnarray}
Explicit solutions for the steady state of $\alpha(t)$ and $\bar\beta_j(t)$ can not be found in general, however, when $\omega_0$ is sufficiently large~\cite{Jie}
\begin{eqnarray}\label{cond0}
\kappa, \ g \abs{\EE_\pm}/\omega_0\ll\omega_0
\end{eqnarray}
it is a valid approximation to expand the solution for $\alpha(t)$ and $\bar\beta_j(t)$ at the lowest relevant order in $g$, and simple analytical expressions can be found. In detail
\begin{eqnarray}
\alpha(t)&=&\alpha_+\ee^{\ii\omega_0t}+\alpha_-\ee^{-\ii\omega_0t}+o(g^2)
\end{eqnarray}
with
\begin{eqnarray}
\alpha_\pm&=&\frac{-\ii\,\EE_\pm}{\kappa\pm\ii\,\omega_0}\ ,
\end{eqnarray}
and
\begin{eqnarray}
\bar\beta_0(t)\simeq \beta_0^+\ee^{2\ii\omega_0t}+\beta_0^-\ee^{-2\ii\omega_0t}+o(g^3)
\end{eqnarray}
with
\begin{eqnarray}
\beta_0^\pm&=&-\ii\,g\, \alpha_\pm\,\alpha_\mp^*\  \vv_0^T\pq{\WW+\ii\,(1\pm 2)\,\omega_0\,\id}^{-1}\vv_0 \ .
\end{eqnarray}

Finally, in the interaction picture with respect to the Hamiltonian
$H_0=\hbar\,\omega_0\sum_{j=-N}^N b_j\da b_j$, the transformed density matrix $\tilde\rho=\ee^{\ii H_0 t/\hbar}\rho' \ee^{-\ii H_0 t/\hbar}$ fullfils a master equation of the form 
\begin{eqnarray}
\dot{\tilde\rho}=-\frac{\ii}{\hbar}\pq{H+\tilde H(t),\tilde\rho}+\pt{\LL_{a}+\LL_{diss}}\tilde\rho\ ,
\end{eqnarray}
where the time independent part of the system Hamiltonian $H$ is equal to Eq.~\rp{H}
with the linearized coupling strength defined as
\begin{eqnarray}\label{Gpm1}
G_\pm=g\,\alpha_\pm\ ,
\end{eqnarray}
while the residual time-dependent part is given by
\begin{eqnarray}
\tilde H(t)&=&
2\,\hbar\,g\,{\rm Re}\pg{\bar\beta_0(t)}\,a\da a
\nn\\&&
+\hbar\sum_{\xi=\pm1}\, \sum_{j=1}^N\ \eta_j\pq{b\da_{\xi\pt{j-1}}b\da_{\xi j}\ee^{2\ii\omega_0 t} + b_{\xi\pt{j-1}}b_{\xi j}\ee^{-2\ii\omega_0 t}   }
\nn\\&&
+\hbar\,a\da\pq{G_-\ee^{-2\ii\omega_0 t}b_0 +G_+\ee^{2\ii\omega_0 t}b_0\da}
\nn\\&&
+\hbar\,a\pq{G_-^*\ee^{2\ii\omega_0 t}b_0\da+G_+^*\ee^{-2\ii\omega_0 t} b_0}\ .
\end{eqnarray}
The time-independent part is the one responsible for the squeezing dynamics as discussed in the main text, while $\tilde H(t)$ accounts for spurious processes which tend to degrade the squeezing dynamics.
When, consistently with Eq.~\rp{cond0}, $\abs{G_\pm},g\abs{\bar\beta_0(t)},\abs{\eta_j}\ll\omega_0 $, the time dependent term can be neglected and we find Eq.~\rp{Meq1}.

\subsection{Optomechanics II: Simulation with an array of optical cavities}
\label{OCchain}

A second possible implementation, complementary to the one just discussed, consists in the realization of the chain in terms of optical cavities, while the squeezed reservoir is engineered by optomechanical interactions between an additional mechanical resonator and the central cavity which in turn is driven by a bichromatic field as in~\cite{Kronwald14}.
 Differently from the conditions identified above, here the steady state entangled dynamics is observed in a different  optomechanical parameter regime corresponding to large mechanical dissipation.

Analogously to the derivation of the previous section, we find that also this system can be described by a model similar to the one introduced in Eq.~\rp{Meq1}
\begin{eqnarray}\label{Meq2}
\dot{\rho}=-\frac{\ii}{\hbar}\pq{H,\rho}+\pq{\LL_{a}\al{n_a}+\LL_{diss}\al{n_{Tj}=0}}\rho\ ,
\end{eqnarray}
with $H$ formally equal to Eq.~\rp{H}, but where now $b_j$ and $b_j\da$ are the operators for the cavity fields realizing the chain of oscillators, while $a$ and $a\da$ are the operators for the mechanical resonator. Furthermore the linearized coupling strength $G_\pm$ have a different dependence on the system parameters and are defined as 
\begin{eqnarray}\label{Gpm2}
G_-=g\,\beta_0^-\ , \ \ \ \ \ \ \ \ \ \ G_+=g\,{\beta_0^+}^*\ ,
\end{eqnarray}
with
\begin{eqnarray}
\beta_0^\pm&=&-\ii\,\EE_\pm\ \vv_0^T\pq{\WW\pm\ii\,\omega_m\,\id}^{-1}\vv_0
\end{eqnarray}
where $\vv_0$ and $\WW$ are defined in Eqs.~\rp{vbvv0} and \rp{WW} respectively.
We note that also the dissipative parts are slightly modified.  The term $\LL_{diss}\al{n_{Tj}=0}$ is equal to Eq.~\rp{LLdiss} with $n_{Tj}=0$, and describes the disspation of the optical cavities at rates $\gamma_j$, while $\LL_{a}\al{n_a}=\kappa(n_a+1)\DD[a,a\da]+\kappa\,n_a\DD[a\da,a]$ describes the dissipation of the mechanical resonator at rate $\kappa$ in a thermal bath with $n_{a}$ average excitations.

In this case if the mechanical dissipation $\kappa$ is sufficiently large then the mechanical resonator can be adiabatically eliminated and the resulting dynamics is described, by a model equal to Eq.~\rp{Meq0}, but where now
\begin{eqnarray}\label{effpar2}
\Gamma&=&\frac{\abs{G_+}^2-\abs{G_-}^2}{\kappa}
\nn\\
\bar n&=&n_a+(2\,n_a+1)\frac{\abs{G_-}^2}{\abs{G_+}^2-\abs{G_-}^2}
\nn\\
\bar m&=&-(2\,n_a+1)\frac{G_+^*\,G_-}{\abs{G_+}^2-\abs{G_-}^2}\ .
\end{eqnarray}

\end{document}